\documentclass[manuscript]{aastex} \usepackage{emulateapj5}

\usepackage{natbib} \usepackage{epsfig} 

\begin{document}

\title{On the Dayside Thermal Emission of Hot Jupiters}

\author{S. Seager\footnote{Department of Terrestrial Magnetism,
Carnegie Institution of Washington, 5241 Broad Branch Rd. NW,
Washington, DC 20015}, L. J. Richardson\footnote{NASA/Goddard Space
Flight Center, Exoplanets and Stellar Astrophysics Laboratory, Mail
Code 667, Greenbelt, MD 20771}$^,$\footnote{Visiting Astronomer at the
Infrared Telescope Facility}, B. M. S. Hansen\footnote{Department of
Physics \& Astronomy, Box 951547, 475 Portola Plaza, University of
California Los Angeles, Los Angeles, CA 90095-1547},
K. Menou\footnote{Department of Astronomy, Columbia University, 1328
Pupin Hall, 550 West 120th Street, New York, NY 10027},
J. Y-K. Cho$^1$, D. Deming\footnote{NASA/Goddard Space Flight Center,
Planetary Systems Laboratory, Code 693, Greenbelt, MD 20771}$^{,3}$}

\begin{abstract}

We discuss atmosphere models of HD209458b in light of the recent
day-side flux measurement of HD209458b's secondary eclipse by
Spitzer-MIPS at 24~$\mu$m. In addition, we present a revised secondary
eclipse IRTF upper limit at 2.2~$\mu$m which places a stringent
constraint on the adjacent H$_2$O absorption band depths. These two
measurements are complementary because they are both shaped by H$_2$O
absorption and because the former is on the Wien tail of the planet's
thermal emission spectrum and the latter is near the thermal emission
peak. A wide range of models fit the observational data, confirming
our basic understanding of hot Jupiter atmospheric physics.  Although
a range of models are viable, some models at the hot and cold end of
the plausible temperature range can be ruled out. One class of
previously unconsidered hot Jupiter atmospheric models that fit the
data are those with C/O $\gtrsim$1 (as Jupiter may have), which have a
significant paucity of H$_2$O compared to solar abundance models with
C/O = 0.5.  The models indicate that HD209458b is in a situation
intermediate between pure in~situ reradiation and very efficient
redistribution of heat; one which will require a careful treatment of
atmospheric circulation.  We discuss how future wavelength-dependent
and phase-dependent observations will further constrain the
atmospheric circulation regime.  In the shorter term, additional
planned measurements for HD209458b, especially Spitzer IRAC
photometry, should lift many of the model degeneracies.
Multiwavelength IR observations constrain the atmospheric structure
and circulation properties of hot Jupiters and thus open a new chapter
in quantitative extrasolar planetology.

\end{abstract}

\section{Introduction}

The star HD209458 is the brightest known ($V=7.5$) to host a
short-period giant planet which transits the face of the
star \citep{char2000, henr2000}.  Recently \citet{demi2005b} detected
the secondary eclipse of the planet by the star in this system, at a
wavelength of 24$\mu$m using the MIPS instrument on the Spitzer Space
Telescope. The measured flux decrement during secondary eclipse
provides a direct measurement of the planetary day side thermal
emission.  Together with the recent Spitzer IRAC secondary eclipse
detection of TrES-1 \citep{char2005}, this marks the first measurement
of the thermal emission from a known extrasolar planet.

In this paper we examine what constraints can be placed on the
planetary properties by this historic measurement. While our principal
constraint is the measured 24~micron flux ($F_{24\mu{\rm m}}$) there
are several additional observational data points useful for the study
of the HD209458b atmosphere.  The most relevant additional constraint
for the planetary day side flux is an upper limit on the H$_2$O
absorption band depths on either side of the 2.2~$\mu$m continuum flux
peak ($\Delta F_{2.2\mu{\rm m}}$), reported by \citet{rich2003}.
Two data points from transmission spectra of the lower
atmosphere---which probe the planetary limb---are relevant: the Na
resonance doublet detection at 0.584~$\mu$m \citep{char2002} and an
upper limit on CO at 2~$\mu$m \citep{demi2005a}. With these data, we
can begin to constrain atmosphere models of the extrasolar planet
HD209458b.

The new Spitzer data is consistent with a circular orbit for
HD209458b, based on the equal time spacing between primary and
secondary eclipses \citep{demi2005b}. In addition, the
latest radial velocity determination of the eccentricity is
essentially zero: $e=0.013 \pm 0.009$ \citep{laug2005}.  With a
circular orbit, the anomalously large radius of HD209458b is not
explained by an interacting companion \citep{bode2003, demi2005b}.  While
atmospheric processes have also been hypothesized to play a role in
the radius evolution of HD209458b \citep{guil2002}, to
first-order, the atmospheric circulation regime on the $\sim$ 3-day
period hot Jupiters with radii closer to Jupiter's is not expected to
differ much \citep{meno2003}.  The $F_{24 \mu{\rm m}}$ also
indicates that the planet HD209458b is in synchronous rotation with
its orbit, presenting the same face to the star at all times. Because
the timescale for synchronization is much smaller than the timescale
for orbital circularization \citep{gold1966}, the tidally-locked
assumption is likely valid.  The notion of a permanent day side has
important consequences for model atmospheres and data interpretation
because it means only one side of the planet is being heated by the
star.

We begin by describing the HD209458b $F_{24 \mu{\rm m}}$ measurement
and $\Delta F_{2.2 \mu{\rm m}}$ upper limit in \S2.  In \S3 we
describe the radiative transfer models and their uncertainties. In \S4
we present a comparison of the models with the HD209458b data as well
as a discussion of the relation between the observations of TrES-1 and
HD209458b.  In \S5 we consider the broader implications and near-term
prospects for both theory (\S5.1) and observation (\S5.2).  We
conclude in \S6 by describing how specific upcoming observations of
HD209458b will enable a more definitive atmospheric characterization.

\section{HD209458b Secondary Eclipse Data}

In this paper we focus on the thermal emission data.  Both $F_{24
\mu{\rm m}}$ and $\Delta F_{2.2 \mu{\rm m}}$ are secondary eclipse
measurements which probe the planetary day side
thermal emission. 

\subsection{The Spitzer 24$\mu$m Flux}

The HD209458b secondary eclipse measured by Spitzer yields a
planet/star flux ratio $F_{24 \mu{\rm m}}/F_* = 0.00260 \pm 0.00046$,
a 5.6$\sigma$ result.  During secondary eclipse the planet is
obscured and only starlight is present. The resulting flux decrement
yields the ratio of planet to stellar flux.  Multiplying this by the
measured stellar flux at this wavelength~($21.17 \pm 0.11$
mJy), the planetary flux is thus measured to be $F_{24\mu{\rm m}} = 55
\pm 10~\mu$Jy.

\subsection{The IRTF 2.2 $\mu$m Constraint}

Hot Jupiter atmosphere models generically predict a peak in the
continuum near 2.2~$\mu$m due to the combined influence of water and
CO absorption (with CO a relatively weak component at 2.3~$\mu$m).
\citet{rich2003} used the method of ``occultation spectroscopy'' to
search for this signature during the HD209458b secondary eclipse using
the NASA Infrared Telescope Facility (IRTF) SpeX \citep{rayn2003}
spectral data from 1.9 to 4.2 microns at a spectral resolution of
$\Delta \lambda / \lambda = 1500$. We have revisited the data to
improve the upper limit.

The data do not provide an absolute upper limit to the 2.2$\mu$m flux
itself. The differential nature of the observations means that only a
relative measurement among the data points in the specified spectral
range is possible.  We therefore calculate the upper limit to the flux
\emph{difference}, or band depth, at the location in the spectrum
where an absorption band is expected. Note that the data points are
expressed in terms of the star-planet flux ratio.  We use three data
bins: a central bin where the flux peak is predicted to reside
(2.09--2.31$\mu$m), and one bin on either side of the flux peak bin
where an absorption band trough should appear (1.955--2.09$\mu$m and
2.31--2.52$\mu$m).  The difference between the flux peak bin and the
absorption trough bins are $2.02 \times 10^{-5} \pm 1.25 \times
10^{-4}$ and $9.78 \times 10^{-6} \pm 1.52 \times 10^{-4}$
respectively.  We average these two values to obtain an average band
depth of $1.5 \times 10^{-5} \pm 9.8 \times 10^{-5}$.  Since the
result is essentially zero, we take the error on the mean $\sim 1
\times 10^{-4}$ to be the upper limit on the band depth on either
side of the 2.2~$\mu$m flux peak at the $1 \sigma$ level.

We convert this contrast ratio limit into a flux limit of $\Delta
F_{2.2 \mu{\rm m}} = 200 \mu$Jy (see Figure~\ref{fig:spectra}), using
a Kurucz model atmosphere for the star \citep{kuru1992} and the
HD209458b planet-star area ratio \citep[using $R_p=1.347 \pm 0.060
R_J$ and $R_* = 1.146 \pm 0.050 R_{\odot}$ from][]{brow2001}.  To
summarize, the $\Delta F_{2.2 \mu{\rm m}}$ constraint means that the
band depths from the continuum peak to the absorption trough must be
$<200 \mu$Jy. This proves to be a useful constraint and is discussed
further in relation to model atmospheres in \S\ref{sec-models}.

\section{HD209458b Atmosphere Models}

\label{sec-models}

Model atmosphere computations are required for an interpretation of
spectral data because the planetary fluxes at different wavelengths
are shaped by a variety of physical processes. Measurements of the
flux level at only one or two wavelengths can be very misleading
if not interpreted in the appropriate model context.

\subsection{Background and Caveats}

Model spectra were computed with a 1D plane parallel radiative
transfer code. This code solves three equations: the equation of
radiative transfer, the equation of radiative equilibrium, and
hydrostatic equilibrium to derive three unknowns: temperature (as a
function of altitude), pressure (as a function of altitude) and the
radiation field (as a function of altitude and wavelength). The
boundary conditions are the stellar radiation at the top of the
atmosphere, and the interior entropy at the bottom of the
atmosphere. The model outputs are temperature, pressure, and radiation
field (from which the emergent flux can be computed). From the
wavelength-integrated thermal emission flux ($F$), $T_{eq}$ can be
calculated by $(F/ \sigma_R)^{1/4}$ where $\sigma_R$ is the radiation
constant, and the geometric albedo\footnote{The Bond albedo (the total
radiation reflected in all directions compared to the total incident
radiation from the star) cannot be computed in these 1D models, and is
difficult to derive from the computed geometric albedo (reflected
radiation relative to that from a flat Lambertian surface of the same
cross-sectional area as the planet). On most solar system planets the
Bond albedo ranges from 10\% to 40\% less than the geometric
albedo. See \citet{cham1987} for more quantitative albedo
definitions.} can be computed by ratioing the visible-wavelength
emergent flux to the incident stellar flux.

Although the physics of this 1D model is straightforward, it is
necessary to choose several input parameters, which thus constitute
uncertainties with various levels of impact.  The interior entropy is
unknown, although it can be estimated from evolutionary models.
Opacities govern the absorption and reemission of radiation and so
drive the entire radiative transfer problem, determining the vertical
temperature-pressure profile and the planetary flux.  The choices of
metallicity, which atomic and molecular species to include, whether to
include only equilibrium or non-equilibrium chemistry as well, all
affect the opacities and are thus input parameters.  Clouds are the
most serious uncertainty in terms of opacities. They form at a the
saturation vapor pressure and (for solar system planets) can extend 1
to 2 pressure scale heights above this ``cloud base''.  Dozens of high
temperature condensates could potentially exist; only a few are
suspected to have a significant magnitude of opacity. The condensate
opacity is controlled by particle size distribution and index of
refraction, as well as particle shape. Although cloud models have been
used by some researchers to compute the particle size distribution and
cloud vertical extent for spherical particles \citep{acke2001,
coop2003}, they still have as a free parameter the type of condensate
and the amount of condensed material in the cloud (e.g., by specifying
the efficiency of sedimentation). All extrasolar planet atmospheres
models in the literature consider horizontally homogeneous clouds
(i.e., covering the entire planet).

Finally, for 1D models we must also make an assumption of how the
incident stellar radiation is redistributed across the surface of the
planet.  We use the parameter $f$ (described in \S\ref{sec-teq}) as
the proxy for atmospheric circulation: $f=1$ if the absorbed stellar
radiation is redistributed evenly throughout the planet's atmosphere
(e.g., due to strong winds rapidly redistributing the heat) and $f=2$
if only the heated day side reradiates the energy (for the latter
$T_{eq}$ is for the day side only). The implementation in the models
is achieved by multiplying the incoming stellar radiation\footnote{The
planet intercepts stellar radiation in a cross sectional area of $\pi
R_p^2$ and reradiates the energy into either $2 \pi R_p^2$ ($f=2$) or
$4 \pi R_p^2$ ($f=1$).} by a factor of $f/4$.

\subsection{Model Choices}

In terms of choices for the models presented here, we use: a Kurucz
model atmosphere for the stellar radiation \citep{kuru1992}; solar
abundances (HD209458A is close to solar metallicity with [Fe/H] =
0.04 \citep{gonz2001}; and line opacities of H$_2$O, CH$_4$, CO, Na,
K, and collision-induced absorption opacities of H$_2$-H$_2$
\citep{bory2001} and H$_2$-He \citep{jorg2000}.  We adopt a planetary
interior energy flux\footnote{Although interior models predict an
interior $T_{\rm eff} \sim 100$~K a hotter temperature is required to
explain HD209458b's large radius \citep{guil2002} and motivates our
adopted planetary interior energy flux corresponding to the $T_{\rm
eff}$ of 500~K.}  corresponding to an effective temperature $T_{\rm
eff}=500$~K. A Gibbs Free energy minimization code was used to compute
chemical equilibrium \citep{seag1999}. H$_2$ Rayleigh scattering and
scattering and absorption by MgSiO$_3$ and Fe vertically and
horizontally uniform and homogenous clouds are included with a
specified particle size distribution (a log normal distribution with
particle radius of 2~$\mu$m and $\sigma=0.05$) and vertical extent
(two pressure scale heights above the cloud base) are also included.
Note that for computational efficiency we use a separate, higher
spectral resolution, radiative transfer code with the same input
assumptions to compute the final emergent spectra based on the T/P
profiles.  A more detailed description of the code and opacity
sources can be found in \citet{seag1999} and \citet{seag2000}.
\citep[See][for choices made by other modelers]{barm2001, suda2003,
burr2005, fort2005}.

With all of the above uncertainties there are many choices that lead
to a wide range of models for a given planet. Rather than an
exhaustive search of the model parameter space, we consider three
models (Figure~\ref{fig:spectra}) which span a range of reasonable
possibilities, to interpret the HD209458b $F_{24 \mu{\rm m}}$ and
$\Delta F_{2.2 \mu{\rm m}}$ data. Model 1 is possibly the simplest case one
can consider; an $f=2$, cloud-free model with solar abundances. The
second case, model 2, is a similarly simple model, but now with $f=1$,
so that some energy is redistributed to the night side.  Model 3
illustrates the effects of thick clouds; it is an $f=1$ model with thick,
high clouds given by the parameters described above.

\section{Data Interpretation}

\subsection{Planetary Equilibrium Temperature}
\label{sec-teq}

The planetary equilibrium effective temperature $T_{eq}$ is a
fundamental parameter of the planetary atmosphere and is useful as a
proxy for the global temperature of a planet atmosphere. To first
order $T_{eq}$ tells us how much radiation is being absorbed by the
dayside of the planet.  In this subsection we relate the measured
$F_{24 \mu{\rm m}}$ to $T_{eq}$.

The measured flux $F_{24\mu{\rm m}} = 55 \pm 10 \mu$Jy can be
converted into a 24$\mu$m brightness temperature $T_{24\mu{\rm
m}}=1130\pm 150$~K \citep{demi2005b}. The brightness temperature is
the temperature of a blackbody emitting the flux equivalent at a
specific wavelength and is not necessarily an accurate representation
of the planet's equilibrium effective temperature $T_{eq}$. Only if
the planet radiates as a blackbody is $T_{eq}$=$T_{24\mu{\rm m}}$.

The spectrum of a hot Jupiter, however, is expected to be very
different from a blackbody due primarily to strong H$_2$O absorption
throughout near-IR and mid-IR wavelengths. In particular, in the
mid-IR (which is the spectral region probed by MIPS) H$_2$O has strong
continuous absorption such that there is no true planetary continuum
expected from models at the MIPS wavelength bands, even though the
relatively flat spectrum makes it appear so. In other words, the
H$_2$O vapor absorption in the 24 $\mu$m region substantially
depresses the planetary flux. This implies that if H$_2$O vapor
absorption is present, the true $T_{eq}> 1130$~K.  
Figure~\ref{fig:spectra} shows the theoretical spectrum of our three
HD209458b example models; each of these models match $F_{24\mu{\rm
m}}$ even though their $T_{eq}$ (1700~K, 1420~K, and 1450~K
respectively) are all hotter than $T_{24\mu{\rm m}}$.  We also show
the blackbody spectrum for these temperatures to demonstrate
how significantly $F_{24 \mu{\rm m}}$ is suppressed by H$_2$O
absorption in these models (a flux depression equivalent to a decrease
of $\sim$ 300--600~K in $T_{24\mu{\rm m}}$).

The true equilibrium temperature is regulated by several competing
physical effects---the amount of irradiation received from the host
star, the fraction that is simply reflected rather than absorbed
(governed by the Bond albedo $A_B$), and the amount of energy circulated 
away from the hotter regions by hydrodynamic processes in the atmosphere.
Considering energy balance between the absorbed
and reemitted flux for a planet with zero eccentricity and
ignoring any (likely much smaller) contribution from the planetary
interior,
\begin{equation}
\label{eq:Teq}
T_{eq} = T_{*} \left( \frac{R_*}{2 a} \right)^{1/2} \left[f (1 -
A_B) \right] ^{1/4},
\end{equation}
where $R_*$ is the stellar radius, $T_*$ is the stellar effective
temperature, $a$ the semi-major axis, and $A_B$ the Bond albedo.  Here
$f$ is the proxy for atmospheric circulation, derived by considering
whether the incident stellar radiation is reemitted into
$4\pi$~steradians ($f=1$) if the absorbed stellar radiation is
redistributed evenly throughout the planet's atmosphere (e.g., due to
strong winds rapidly redistributing the heat) or reemitted into $2\pi$
steradians ($f=2$) if only the heated day side reradiates the energy
(for the latter $T_{eq}$ is for the day side
only). Figure~\ref{fig:teq}a shows $T_{eq}$ for the HD209458
parameters \citep[$T_* = 6000$~K, $R_*=1.18R_{\odot}$, and
$a=0.046$~AU;][]{maze2000, cody2002}.

An equilibrium temperature $T_{eq}=1700$~K is at the hot end of the
range for the HD209458 system parameters, as shown in
Figure~\ref{fig:teq}.  If our model 1 with $T_{eq}=1700$~K were the
unique, correct planetary model, this hot $T_{eq}$ would have
significant implications: it is only plausible with both a low $A_B$
and no heat redistribution (i.e., $f=2$).  At the low end of the
plausible temperature range, $T_{eq}=1130$~K is unlikely.  A
$T_{eq}=1130$~K model would require that HD209458b has a very high
Bond albedo ($> 0.65$), and would probably require $f=1$ because
$A_B>0.8$ is difficult to attain. Most of the solar system planets
have $A_B< 0.35$, with the exceptions of Venus ($A_B = 0.75$) and
Pluto ($A_B = 0.4$) \citep{depa2001}.

\subsection{Constraints on Models}

Model 1 (solar metallicity, no clouds, no energy redistribution) is
characterized by strong near-IR H$_2$O vapor absorption bands.  The
H$_2$O absorption also causes a strong depression of $F_{24
\mu{\rm m}}$.  With the strong alkali metal absorption dominating
H$_2$ Rayleigh scattering at visible wavelengths, this model has a low
albedo (geometric albedo = 0.15) and hence high $T_{eq}=1700$~K.  

The Na detection and CO upper limit transmission spectroscopy are both
too weak to match a solar abundance cloud-free chemical equilibrium
model \citep{char2002, demi2005b} and therefore in general provide a
useful model constraint. In this cloudless case the Na could be
transported to the night side and converted to Na$_2$S at the lower
temperatures \citep{guil2002, iro2005}; similarly CO would be
converted to CH$_4$ on the night side, thus also potentially
satisfying the limits from transmission spectroscopy.  It is important
to keep in mind that the transmission spectroscopy probes the
planetary limb; with $f=2$, the day and night side are assumed to be
different temperatures and so the planetary limb could have different
conditions from those computed on the day side, for example a high
cloud at the limb due to the falling temperatures away from the
substellar point. This high cloud would also satisfy the Na and CO limits
from transmission spectroscopy.

We use this $T_{eq}=1700$~K model as an example of a model that is
ruled out by the data. Although this model satisfies $F_{24 \mu{\rm
m}}$ and can satisfy the transmission spectra data, it is ruled out
(at the $5\sigma$ level) by the $\Delta F_{2.2 \mu{\rm m}}$ data. The
strong water vapor absorption bands adjacent to 2.2$\mu$m are in
conflict with the $\Delta F_{2.2 \mu{\rm m}}$ limit.  The $\Delta
F_{2.2 \mu{\rm m}}$ constraint means that the band depths from the
continnuum peak to the absorption trough must be $<200 \mu$Jy, whereas
in this model the band depths are $\sim 1000 \mu$Jy. We used this simple
model in \citet{demi2005b} primarily to illustrate that $T_{eq}$ does
not have to equal $T_{24\mu{\rm m}}$.  The fact that model~1 is ruled
out by our revised 2.2~$\mu$m upper limit shows that $\Delta F_{2.2
\mu{\rm m}}$ is a useful constraint on models.

Model 2 (in which energy is now redistributed to the night side) has a
lower computed $T_{eq} = 1420$~K.  While this model fits $F_{24
\mu{\rm m}}$, it only fits the $\Delta F_{2.2 \mu{\rm m}}$ constraint
at the $3\sigma$ level due to the strong H$_2$O absorption
features. In the context of this $f=1$ cloud-free, solar abundance
model we note that the atmospheric circulation could allow for
substantially varying temperatures at different horizontal locations
\citep[e.g., ][]{show2002, cho2003, coop2005} and so more detailed
models may show partial cloud cover and lower Na abundances from
molecular condensation (which may also better fit the weak Na and CO
transmission spectra).

In Model 3 we allow for formation of thick, absorptive clouds and so,
in contrast to models 1 and 2, it has very weak absorption band
features with a computed $T_{eq}=1450$~K.  The absorption and
remission and scattering from largely grey cloud particles causes the
absorption bands, including the continuous H$_2$O absorption at 24
microns, to be much shallower than in a cloud-free model. The cloud
bases are computed to be at 5 and 10 millibar for MgSiO$_3$ and Fe
respectively, with the cloud tops two pressure scale heights above at
0.8 and 1.3 millibar; this is consistent with both the
lower-than-expected Na and CO transmission spectra.  As seen from
Figure~\ref{fig:spectra} this model fits both the $\Delta F_{2.2
\mu{\rm m}}$ and $F_{24 \mu{\rm m}}$ data.  We emphasize that the
amount of condensates and their opacity strength (controlled by
particle type and size distribution for spherical particles) hugely
affects the overall spectrum; the condensate opacity is competing with
the H$_2$O opacity to determine the water absorption band strengths.
Different choices for the particle type, fraction of species
condensed, cloud vertical scale height, and particle size distribution
will all affect the overall opacity. Just one example is that larger
particle sizes are likely to be more reflective and would not weaken
the absorption bands nearly as much as absorptive particles.

\subsection{Consequences of Model Constraints}

Based on our investigation of models 1 through 3, we can draw some
general conclusions. We showed above that the $T_{eq}=1700$~K model,
under our adopted input assumptions (described in \S4), is ruled out
based on the $\Delta F_{2.2 \mu{\rm m}}$ absorption band depth
constraint. Other hot models are also ruled out by the data. First, a
truly isothermal cloud-free atmosphere would satisfy the $\Delta
F_{2.2 \mu{\rm m}}$ because a vertically isothermal LTE atmosphere
does not have any spectral features---i.e., it is a blackbody
spectrum\footnote{Recall that in LTE with the source function equal to
a Planck function, the solution to the 1D plane parallel radiative
transfer equation is $F(\nu) = 2\pi \int_0^1 I(\tau,\mu,\nu) \mu d\mu
= 2 \pi \int_0^1 \int_0^{\infty} B(T(\tau),\nu) \exp(-\tau/\mu) d\tau
d\mu$, where $T$ is a function of the altitude expressed on an optical
depth scale $\tau$, $\mu=\cos \theta$ where $\theta$ is the angle away
from the normal, $I$ is the intensity, and $\nu$ is frequency. Taking
$B(T(\tau), \nu)$ as a constant out of the integral we see that for an
isothermal atmosphere $F(\nu) = \pi B(T,\nu)$.}.  However, such an
isothermal $T_{eq}=1700$~K model is ruled out by the $F_{24 \mu{\rm
m}}$ data even considering the $3\sigma$ error on $T_{24 \mu{\rm m}}$
which permits a value as high as 1580~K.  Second, most hot models in
between the isothermal atmosphere and the strong vertical temperature
gradient atmosphere are likely ruled out; as the vertical temperature gradient
gets shallower and satisfies $\Delta F_{2.2 \mu{\rm m}}$ with a weaker
absorption band, the 24~$\mu$m model flux gets higher and away from
the measured $F_{24 \mu{\rm m}}$. Third, a hot atmosphere with clouds
that are highly absorptive (low $A_B$), such as Fe clouds, could also
result in a blackbody spectrum (even without an isothermal
atmosphere)---again a hot blackbody or near blackbody spectrum is ruled
out because $T_{24\mu{\rm m}}=1130$~K is lower than resulting values
of $T_{eq}$.

A second category of models that are difficult to fit are those with
$T_{eq}$ at the cold end of the plausible range (equation~(1)
and Figure~\ref{fig:teq}). As described above, for HD209458b to have a
$T_{eq}$ near 1130K, a very high $A_B > 0.65$ is required.  Such a
high $A_B$ could be possible only from a narrow range of cloud
parameters and is not attainable in the absence of clouds---i.e., due
to Rayleigh scattering alone. 

Both of our $f=1$ and $f=2$ cloud-free models have a strong H$_2$O
absorption band on either side of the 2.2~$\mu$m flux peak that do not
fit the $\Delta F_{2.2 \mu{\rm m}}$ constraint at the $1\sigma$
level. What could cause weaker H$_2$O absorption features to fit
$\Delta F_{2.2 \mu{\rm m}}$?  Absorption and remission by cloud
particles weaken the 2.2~$\mu$m absorption feature and are hence
consistent with the data.  A more isothermal atmosphere would also
have weaker absorption features (see above). Decreasing the water
vapor abundance could also serve to weaken the water vapor absorption
features adjacent to 2.2~$\mu$m. This situation is hard to physically
realize in a solar abundance atmosphere. Photodissociation from
stellar UV radiation affects water only in the upper atmosphere, close
to the microbar pressure level \citep{lian2003} where the low H$_2$O
abundance would have little effect on the overall planetary spectrum.

\subsection{Low H$_2$O Abundance in the C/O $\gtrsim$ 1  Regime}

A possible scenario with a very low abundance of water vapor, that
would satisfy $\Delta F_{2.2 \mu{\rm m}}$, is one with atmospheric
abundances of the carbon-to-oxygen ratio C/O $\gtrsim 1$.  This is in
contrast to the solar C/O ratio used in all extrasolar planet models
published so far \citep[0.5 or 0.42,][respectively]{alle2002,
ande1989}.  In addition to being metal-rich overall, Jupiter is likely
enriched in C with C/O = 1.8 \citep{lodd2004}. In addition, Saturn has
less oxygen enrichment than other heavy elements, including carbon
\citep{viss2004}. The enhanced metallicity on Jupiter and the other
solar system planets is thought to come from post-formation
planetesimal accretion and may well be applicable to hot Jupiters;
enriched C compared to O over solar abundances would require pollution
from carbon-rich planetesimals.

A high C/O ratio $\gtrsim1$ can have a large effect on the spectral
signatures of hot planetary atmospheres \citep[see Figure~\ref{fig:co}
and Figure 1 in ][Seager \& Kuchner, in preparation]{kuch2005}.  This
is because at high temperatures the dominant carbon-bearing molecule
is CO, and in equilibrium formation of CO is chemically favored
instead of H$_2$O.  In an extreme case, then, with all of the O tied
up in CO, there will be no H$_2$O vapor present.  In reality,
depending on the temperature-pressure profile, there will be a small
amount of water vapor present, but a few to $>$ eight orders of
magnitude below the CO number density.  This extreme paucity of H$_2$O
is significant, because the hot Jupiter spectra are normally
predominantly shaped by H$_2$O absorption bands. In the C/O $>$ 1 case
the planetary spectrum will instead be dominated by CO absorption and
by CH$_4$ absorption (which forms at higher temperatures than for C/O
$<1$).  (Note that for 0.5 $<$ C/O $<$1, the water vapor will decrease
much less significantly, by a factor of a few to 100
respectively. See \citet{fort2005} for a discussion of the effects of
C/O = 0.7.)  In contrast to the hot Jupiter case, a C/O ratio $\gtrsim
1$ has a much less significant effect on the spectral signatures of
planetary atmospheres cooler than the hot Jupiters (below $ T_{eq}
\sim$ 1000~K). In this case CH$_4$ is the dominant carbon-bearing
molecule regardless of the C/O value; whatever O is present is free to
form H$_2$O.

Any hot Jupiters with C/O $\gtrsim$ 1 should have a very low amount of
water vapor, more so for those at the hot $T_{eq}$ range and those
with hot upper atmospheres.  Figure~\ref{fig:co} shows an estimated
thermal emission spectrum (using a scaled model 1 T/P profile to
reproduce a $T_{eq} = 1600$~K) with (condensate-free) chemical
equilibrium computed for a value\footnote{We chose the value C/O =
1.01 because graphite clouds tend to deplete the gaseous carbon close
to a C/O ratio of 1 \citep{lodd1997}.} of C/O = 1.01.  More careful
modeling is needed to explore the consequences of a carbon-enriched
hot Jupiter atmosphere, particularly for condensate formation which is
affected by the O abundance.  For C/O $>1$ graphite and SiC clouds can
form which are not present in C/O $<1$ models.

\subsection{HD209458b and TReS-1 Comparison}

We now turn briefly to a discussion of the thermal emission detected
from TrES-1, and how it relates to HD209458b day side thermal
emission.  The secondary eclipse detection of TrES-1 \citep{alon2004}
by \citet{char2005} at Spitzer IRAC bands (4.5 and 8 $\mu$m) was
reported at the same time as the HD209458b secondary eclipse
detection. A planetary brightness temperature was reported at each
bandpass: $T_{4.5\mu{\rm m}} = 1010\pm60$~K and $T_{8\mu{\rm m}} =
1230\pm110$~K, with an average $T_{\rm TrES-1} = 1060 \pm 60$~K.  In
this section we consider the TrES-1 data as it relates to HD209458b.

As for HD209458b, $T_{\rm TrES-1}$ does not constrain $A_B$ due to the
unknown atmospheric redistribution of absorbed stellar radiation
\citep[$f$ in equation~$\left( 1 \right)$; c.f.][] {char2005}.  If
$T_{\rm TrES-1} = T_{eq} = 1060$~K \citep[shown in
Figure~\ref{fig:teq}b for the TrES-1 parameters $R_* = 0.85
R_{\odot}$, $T_*=5250$~K, $a = 0.039$~AU;][]{alon2004}, then $0.3 <
A_B < 0.7$, a large range.  Using the same arguments as for HD209458b
(non-blackbody-nature of the spectrum due primarily to H$_2$O
absorption) the $T_{eq}$ is likely to be hotter than 1100K. A very
interesting consequence of TrES-1 stellar temperature and planet
semi-major axis is that {\it if $T_{eq}>1200K$ an $f=1$ model is ruled
out}\footnote{Recall that $T_{eq}$ is for the planetary day side.},
otherwise we would have the unphysical situation $A_B<0$.

We emphasize that HD209458b and TrES-1 are likely to be different
planets in terms of their atmospheric structure.  We caution that
the similar measured brightness temperatures of TrES-1 and HD209458b
are misleading; they are at different wavelengths and probably probe
different atmospheric altitudes/temperatures.  Furthermore, TrES-1 and
HD209458b are theoretically likely to have different $T_{eq}$ due to
the former's intrinsically fainter parent star (K0V ($T_{\rm eff} =
5250$~K) vs. G0V ($T_{\rm eff} = 6000$~K) respectively; see
Figure~\ref{fig:teq}). This temperature difference could have
important consequences in terms of cloud altitude and the CO/CH$_4$
abundance ratio. A cold model for TrES-1 would have more CH$_4$ in the
atmosphere than the hotter HD209458b models, for the same
abundances. The two planets are also different in their radii (for
HD209458b $R_P=1.347 \pm 0.060 R_J$ \citep{brow2001} and for TrES-1
$R_P=1.08_{-0.04}^{+0.18} R_J$ \citep{alon2004}).

Although the two planets are different, it is still instructive to
compare the TrES-1 data to the HD209458b models
(Figure~\ref{fig:spectra}), since the solar abundance models have the
same generic H$_2$O and CO absorption features at 4.5 and 8~$\mu$m.
It is certainly puzzling that the $8\mu$m flux is $\gtrsim$ the
$4.5\mu$m flux, because one generally expects more thermal emission at
shorter wavelengths (see Figure~\ref{fig:spectra}).  The TrES-1 data
constrain the relative flux ratios of a shallow H$_2$O band at $8
\mu$m and a deeper H$_2$O + CO band at 4.5 microns. For most model
fluxes that match the $8\mu$m band the model fluxes are too high in
the $4.5\mu$m band \citep{char2005} but only marginally so considering
the $3\sigma$ errors on the data.

One solution to the IRAC $8\mu$m and $4.5\mu$m data discrepancy
\citep{char2005} is a missing source of opacity at 4.4$\mu$m to cause
a deeper absorption band. This discrepancy is at face value opposite
to the requirement of weak H$_2$O bands for HD209458b from $\Delta
F_{2.2 \mu{\rm m}}$.  Additional H$_2$O opacity would not likely solve
the IRAC measurement discrepancy, because H$_2$O is present in both
observed IRAC bands. An increased CO abundance without an increased
H$_2$O abundance could potentially solve the problem because CO
absorbs in the 4.4 $\mu$m band and not the 8 $\mu$m. A higher CO
number density and lower H$_2$O number density could be realized with
a higher C/O ratio than solar (but not necessarily C/O $>$1), as long
as not too much CH$_4$ is present (which also absorbs in the
8$\micron$ IRAC band and whose number density grows as C/O increases)
(Seager \& Kuchner, in preparation).  See \citet{fort2005} for a
discussion of metallicity and varying C/O ratios for TReS-1 models.  A
second solution to fit the nearly equal fluxes in the IRAC $8\mu$m and
$4.5\mu$m bands is a close-to-vertically isothermal model with weaker
absorption bands. Such a model would have to be a cool model to fit
the data, as illustrated by the blackbody fit in
Figure~\ref{fig:spectra} and in \citet{char2005}.  We note that a
cooler model fit indicates that efficient redistribution of absorbed
stellar radiation (i.e., $f=1$) is likely to be present whereas for a
hotter model it is not (see above).  It is therefore important for
future observations to distinguish between the best fit models: a
hotter model with strong absorption bands or a colder model with weak
absorption bands.

\section{Atmospheric Circulation: Self-Consistent Models and Future Observational Diagnostics}

We now turn from interpreting data, to examining future prospects for
theoretical models. One of the major simplifications in the above
models is the basic assumption about how the absorbed stellar
radiation is redistributed throughout the planet atmosphere, with an
extreme $f=1$ or $f=2$ assumption.  Indeed, this is one of the most
compelling issues to understanding the hot Jupiter data and models
alike. With a permanent day side facing the star from tidal locking
and intense stellar radiation from the very small semi-major axes ($<
0.05$~AU) strong winds are expected to develop to advect the heat away
from the planetary day side \citep[e.g., ][]{show2002, cho2003,
meno2003, coop2005}.  In this section we describe why such
self-consistent coupled radiative transfer/atmospheric circulation
models are required, and discuss the possibility of constraining the
atmospheric circulation regime on HD209458b (and other hot Jupiters by
extension) from wavelength- and time-dependent photospheric signatures
such as $F_{24 \mu{\rm m}}$.

\subsection{Radiative and Advective Regimes}

\label{sec-radadv}

In the simplest terms, the influence of atmospheric circulation on the
planetary atmosphere horizontal temperature-pressure structure can be
discussed as a competition between a typical radiative timescale
($\tau_{\rm rad}$) and a typical advective timescale ($\tau_{\rm
adv}$) \citep[e.g., ][]{show2002, iro2005}.  One can roughly estimate
the relevant timescales from the following relations\footnote{On the
planetary dayside, vertical heat transport by convection is not
important until very deep in the atmosphere because stellar
irradiation imposes a shallow vertical temperature
gradient. Additionally, we note that the definition of the radiative
timescale adopted here is valid only near the photosphere 
\citep[i.e., $\tau \sim 1$; e.g., ][]{iro2005}.}:
\begin{equation}
\tau_{\rm rad} \sim \frac{P}{g}\frac{c_P}{4 \sigma T^3}
\end{equation}
\begin{equation}
\tau_{\rm adv} \sim \frac{R_P}{U},
\end{equation}
where $P$ is the pressure, $g$ is the surface gravity, $c_P$ is the
heat capacity, and $U$ is the characteristic wind speed ({\it a priori}
unknown).  We are interested in the ratio of these quantities at the
location where most of the incoming energy is deposited.  This is
determined largely by the opacity of the atmosphere at visible
wavelengths.

One can ask: is the bulk of the stellar radiation absorbed in the
radiative regime $\tau_{\rm rad}/\tau_{\rm adv} \ll 1$, or in the
advective regime $\tau_{\rm rad}/\tau_{\rm adv} \gg 1$?  If the
atmosphere is in the radiative regime, the bulk of the absorbed energy
would be reemitted before being advected to the night side. In this
case (which corresponds to our $f=2$ model) we would expect a strongly
phase-dependent horizontal temperature gradient centered on the
substellar point, and a strong day-night temperature difference.  If,
on the contrary, the bulk of the stellar energy is absorbed in the
advective regime, then atmospheric circulation \citep{show2002,
cho2003, coop2005} is directly required to understand the horizontal heat
transfer. The horizontal temperature field should then be much more
uniform (corresponding to our $f=1$ model) because heat is transported
and redistributed efficiently over the entire planet.

The main issue in understanding in which regime we are is the unknown
wind speed $U$.  If we consider $U=1000$~m/s throughout the atmosphere
(for illustration purposes) we can investigate the regime where the
bulk of stellar radiation is absorbed in our models.  For our $f=2$
cloud-free model~1, (Figure~\ref{fig:atmcirc}a) most of the stellar
radiation is absorbed at or above 1 bar. This is right on the boundary
near $\tau_{\rm rad}/\tau_{\rm adv} \sim 1$. For a smaller $U$, we
would be in the radiative regime, and for a larger $U$ in the
advective regime (since $\tau_{\rm rad}/\tau_{\rm adv} \propto U$). On
the other hand, in our $f=1$ model~3 with very thick clouds, much of
the stellar radiation is absorbed much higher in the atmosphere,
around 5 mbar (Figure~\ref{fig:atmcirc}a). In this case, the stellar
energy is absorbed well into the radiative regime.  One notices
immediately that this conclusion contradicts the basic assumption of
the $f=1$ model, namely that energy is efficiently redistributed! This
ambiguity is probably not fatal since the model improvements
discussed below will alleviate this.

Considering windspeeds different from $U=1000$~m/s we see that,
according to the simple timescale comparison, efficient transport of
heat in atmospheric regions with very small $\tau_{\rm rad}$ requires
a supersonic speed \citep{coop2005}\footnote{The sound speed is $\sim
2,500$~m/s at $\sim 1500$~K.}.  While there is no {\it a priori}
reason to disregard supersonic winds from first principles, such a
situation would be remarkable because {\it all} observed wind speeds
on solar system planets are well in the subsonic regime. This unusual
situation could be attributed to the extremely close proximity to the
star and correspondingly higher temperatures and shorter $\tau_{rad}$
values for hot Jupiters (see equation~(2))\footnote{One should
exercise caution, however, since Jupiter, the giant planet with the
weakest cloud-deck winds in our solar system, is also the closest to
the Sun \citep[See Table~1 in][]{meno2003}.}  From a theoretical viewpoint, if circulation with
large-scale supersonic winds exist, it poses a very interesting
problem for flow adjustment dynamics. Supersonic winds are not
required, however, to redistribute the energy in the short $\tau_{\rm
rad}$ regime; other mechanisms such as subsonic winds/advection,
atmospheric waves, or horizontal radiative transfer can cause energy
to be transported or to diffuse away from the hot planetary day side.

All of the above indicate the need for coupled radiative-atmospheric
circulation models. In particular we need to address the unknown value
of $U$, the fact that the bulk of the stellar radiation is absorbed
near the boundary between the radiative and advective regimes and the
apparent contradictions in 1D model assumptions for heat
redistribution. Furthermore, the above simple $\tau_{\rm
rad}/\tau_{\rm adv}$ timescale argument cannot replace combined
radiative transfer-atmospheric circulation calculations because these
two atmospheric timescales are actually non-linearly coupled:
atmospheric winds are driven by pressure gradients due to uneven
radiative heating in this context of strong insolation. The value of
$U$ in equation~(3) depends on $T$ (wind forcing) while $T$ itself
depends on $U$ because of heat advection by the winds.

What do we mean by a coupled radiative transfer/atmospheric
circulation model? The 1D emergent spectra (\S4) depend on the
vertical temperature gradient.  This temperature gradient, in fact,
depends on the atmospheric circulation to determine the temperatures
at various altitudes.  Further, in the atmospheric circulation picture a
1D model is not valid, and a computation of different vertical (i.e.,
radial) T/P profiles away from the substellar point, followed by the
subsequent hemispherical integration of the emergent spectra is
necessary.  As such, no truly self-consistent models yet exist in the
literature. As more data accrues, such consistent models
will become necessary.

\subsection{Wavelength-Dependent, Phase-Dependent Observational Diagnostics}

Infrared observations of hot Jupiters can provide data to help
constrain the atmospheric circulation via measured horizontal
temperature gradients.  These observations include: day-night
temperature differences measured for transiting hot Jupiters; IR phase
curves as a hot Jupiter planet orbits its parent star; and the rate of
secondary eclipse ingress and egress which will differ if the
planetary dayside is nonuniform in temperature.

Assuming that we can distinguish observationally between strong
horizontal temperature gradients and a uniform horizontal temperature,
what can be told about atmospheric circulation?  In this
section we describe how wavelength-dependent, phase-dependent
observations can potentially be used to constrain the planet's
atmospheric circulation regime.

The key point is that the altitude of thermal emission is wavelength
dependent, due to wavelength-dependent opacities. This means that at
different wavelengths we can potentially ``see'' altitudes with
different ratios of the radiative to advective
timescales. Additionally, the stellar radiation is absorbed at visible
wavelengths, heats the atmosphere, and is reemitted at infrared
wavelengths. The altitude where visible radiation is absorbed compared
to the altitude where thermal reemission occurs are only weakly
coupled in the absence of clouds: the chemistry of molecular and
atomic species (Na, K, H$_2$-H$_2$ collision-induced opacities) which
control the absorption of stellar irradiation at visible wavelengths
is not directly coupled to the chemistry of molecules that control
thermal reemission at IR wavelengths (predominantly H$_2$O).  The IR
spectrum, therefore, allows us to probe different atmospheric heights,
independently of the altitude where the bulk of stellar radiation is
absorbed.  The altitude dependence of thermal emission is shown
explicitly in Figure~\ref{fig:atmcirc}a for the models previously
considered. Thermal emission at 24 $\mu$m occurs relatively high in
the planetary atmosphere ($P \sim 10$~mbar). Wavelengths at the
continuum between H$_2$O bands (in particular, in the 1--2$\mu$m range
and at the 4$\mu$m continuum peak) probe much deeper altitudes closer
to 1 bar. Thus, we could potentially probe different regimes of
atmospheric circulation in a single atmosphere using multi-wavelength
observations.

We consider two extreme cases, where the radiation is absorbed deep in
the atmosphere ($\sim$ 1 bar) or high in the atmosphere (near millibar
pressures).

Building on our discussion in the previous section, if the stellar
radiation is absorbed deep in the planetary atmosphere, at pressures
$> 1$~bar, we are likely to be in the advective regime.  In this
regime, redistribution of the absorbed stellar energy by the
atmospheric circulation will be significant. Higher up in the
atmosphere, the temperature field will be determined by a combination
of horizontal advection, vertical radiative transport through the
various atmospheric layers, and horizontal and vertical wave
propagation, but the initial redistribution of heat at the absorption
levels should have a significant impact on the overall temperature
structure in the entire atmosphere.  Different predictions for the
temperature field expected in this advective regime exist \citep[][Cho
et al., in preparation]{show2002, coop2005, cho2003} and they should
ultimately be testable with observations in the future once they are
combined with detailed spectral predictions.

If the stellar radiation is absorbed at high atmospheric altitudes, in
the radiative regime, different observational signatures are
expected. In this case the energy is expected to be reemitted before
it can be advected to the planetary night side. One expects a strong
dayside horizontal temperature gradient, centered on the substellar
point.  Such a strong temperature gradient could be detected from the
rate of ingress (and egress) which would be different from the uniform
horizontal temperature case.  An additional observational diagnostic
is a very strong day-night temperature difference, which may
be detectable by accurate flux differencing from near primary and
secondary eclipses.

If we believe the stellar irradiation is absorbed very high in the
atmosphere where the radiative timescale is short, and no strong
temperature gradients are inferred (particularly none centered on the
substellar point) then an interesting possibility exists: very high
winds which may even be supersonic. However, other processes may also
mitigate the extreme temperature gradient (see \S\ref{sec-radadv}). An
observational diagnostic to differentiate between supersonic winds and
subsonic processes is Doppler broadening of atomic lines which could
potentially be detected in future observations of primary transit
transmission spectral lines \citep{brow2001}.

In short, the prospect of multiple wavelength, phase-dependent observations
offers the possibility of building a picture of the three dimensional
temperature field on the surface of planets such as HD209458b and of
constraining the nature of the atmospheric circulation.

\section{Summary and Near Future Prospects}

\label{sec-discussion}

\begin{figure}
\begin{center}
\epsfig{file=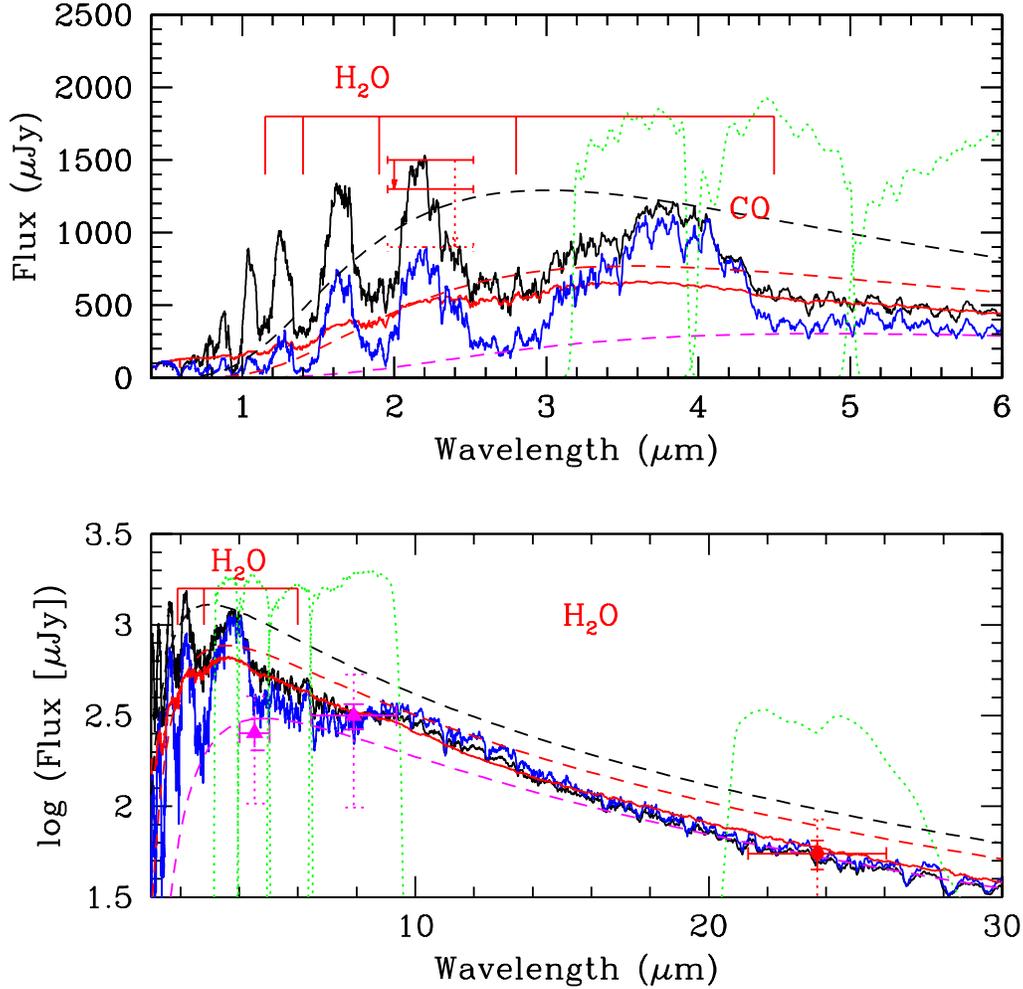, width=14cm, height=14cm}
\end{center}
\caption{Theoretical spectra of HD209458b.  The upper panel shows the
1 to 6 micron window which is expected to contain many H$_2$O
absorption features, and the lower panel shows the spectrum from 1 to
30 microns with log flux on the y axis.  Black curve: model 1,
$T_{eq}=1700$~K, $f=2$ cloud-free model; blue curve: model 2,
$T_{eq}=1450$~K, $f=1$ cloud-free model; red curve: model 3,
$T_{eq}=1420$~K, $f=1$ cloudy model.  Note how the red curve has very
weak absorption bands. The dashed curves show blackbody spectra with
$T=1700$~K (black curve) and $T=1430$~K (red curve); HD209458b data
points (red solid error bars 1$\sigma$, red dotted error bars
3$\sigma$) are shown as follows. IRTF $\Delta F_{2.2 \mu{\rm m}}$
upper limit: the absorption band depth on either side of the 2.2
$\mu$m flux peak must be smaller than the vertical difference between
the horizontal bars shown (indicated by the arrows). Because $\Delta
F_{2.2 \mu{\rm m}}$ is a relative measurement, these horizontal bars
can be moved along the y axis.  Spitzer MIPS 24$\mu$m point is
shown. A hot blackbody is ruled out by the MIPS data, while model 1
($T_{eq}=1700$~K) is ruled out by $\Delta F_{2.2 \mu{\rm m}}$.
Spitzer IRAC TrES-1 4.5$\mu$m and 8$\mu$m data are scaled for
comparison with HD209458b (magenta triangles and error bars). A
$T=1060$~K blackbody (magenta dashed curve) is shown that fits the
IRAC data. The renormalized Spitzer IRAC and MIPS band passes are
shown as green dotted lines.
\label{fig:spectra} }
\end{figure}

\begin{figure}
\begin{center}
\epsfig{file=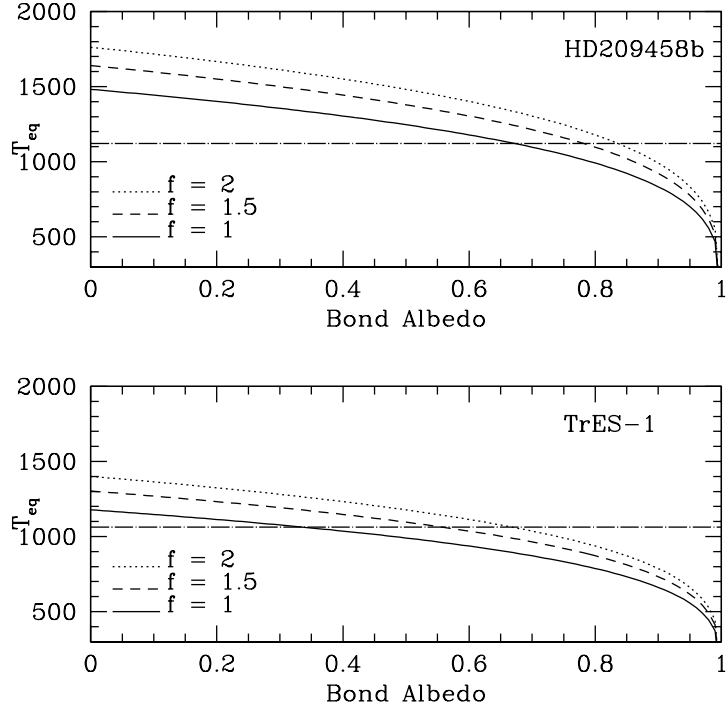, width=10cm, height=10cm}
\end{center}
\caption{The dayside $T_{eq}$ as a function of $A_B$ for different
values of $f$ (see equation~1). Top panel: $T_{eq}$ for the HD209458
system parameters. The 24 $\mu$m brightness temperature $T_{24} =
1130$~K is shown (dash-dot line); if $T_{eq} = T_{24}$ the planet must
have a very high Bond albedo. Bottom panel: $T_{eq}$ for the TrES-1 system
parameters. The Spitzer IRAC measured brightness temperature of 1050~K
is shown (dash-dot line). If $T_{eq}>1200$~K, $f=1$ models are ruled
out. Although the measured brightness temperatures for HD209458b and
TrES-1 are similar (albeit measured at different wavelengths), values
of $T_{eq}$ as a function of $A_B$ show that the viable range  of planet
equilibrium temperature are quite different.
\label{fig:teq}}
\end{figure}

\begin{figure}
\begin{center}
\epsfig{file=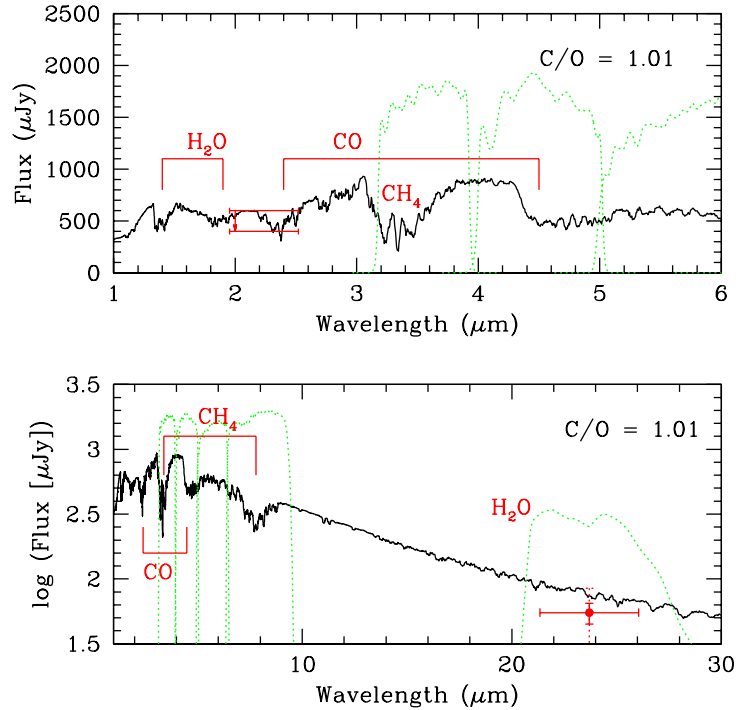, width=10cm, height=10cm}
\end{center}
\caption{Thermal emission spectrum for HD209458b with C/O = 1.01 (and
other elements in solar abundance). This chemical regime results in
much reduced water abundance and increased CH$_4$ opacity compared to
the solar abundance spectra shown in Figure~1.  The $\Delta
F_{2.2 \mu{\rm m}}$ constraint and $F_{24 \mu{\rm m}}$ measurement are
shown, as are the IRAC and MIPS bandpasses (dotted lines).
\label{fig:co}}
\end{figure}

\begin{figure}
\begin{center}
\epsfig{file=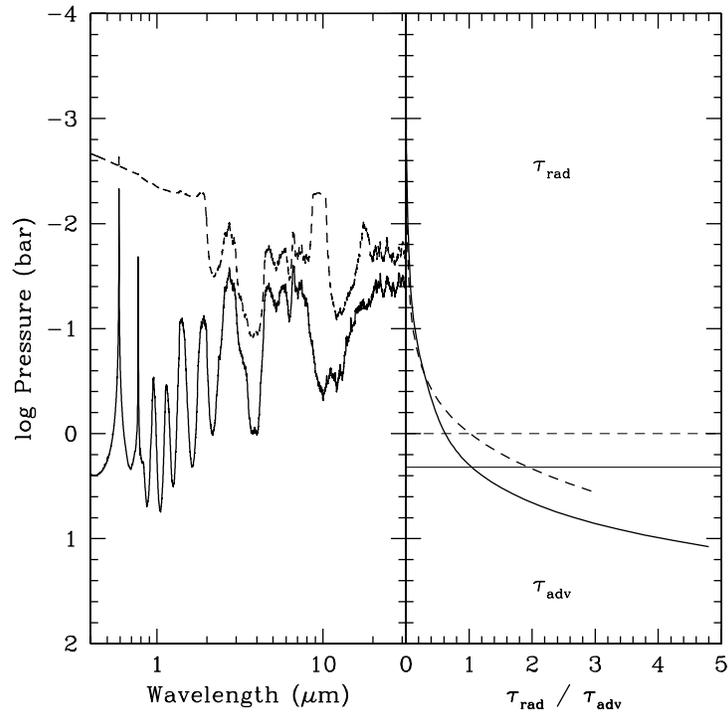, width=10cm, height=10cm}
\end{center}
\caption{The photosphere and atmospheric circulation. Left panel:
Pressure (as a proxy for altitude) at the planetary photosphere as a
function of wavelength.  (We define the photosphere at an optical
depth of 2/3.) The f=2 cloud-free model (model 1) is shown as the
solid curve, the f=1 cloudy model (model 3) is shown as the dashed
one.  Right panel: the altitude dependence of the ratio of the
radiative to advective timescales $\tau_{\rm rad}/\tau_{\rm adv}$. A
windspeed $U$ of 1000 m/s was adopted for illustration; the ratio
scales linearly with $U$ so that other values can be considered. See
text for discussion.
\label{fig:atmcirc}}
\end{figure}

We have shown that, to first order, 1D radiative transfer models can
fit the HD209458b thermal emission data at 24 microns \citep[see
also][]{burr2004, burr2005, fort2005}.  This confirms not only that
the models work, but also that hot Jupiters are indeed hot and likely
heated externally by their parent stars.  If we conservatively adopt
the $3\sigma$ error bars (Figure~\ref{fig:spectra}), a wide range of
models fit the data and we are far from a unique interpretation of the
atmosphere of HD209458b.  We have described a previously unconsidered
set of models that can fit the data, hot Jupiter atmosphere models
with C/O $\gtrsim$1 which have a significantly different chemical
equilibrium than the C/O = 0.5 solar abundance models, in particular
an extremely low abundance of water vapor.  Despite the viability of a
wide range of models the data do allow us to begin to rule out some
models, specifically hot, cloud-free models and models at the cold end
of the plausible $T_{eq}$ range. Because the very hottest models are
ruled out---models only possible with $f=2$---a situation intermediate
between pure radiative equilibrium and very efficient redistribution
of heat is likely, which will require a careful treatment of
radiative-advective processes.

Observations planned for the next year will help towards a more
definitive characterization of HD209458b.  These observations are
listed in Table~1.  The most useful planned measurements will be the
Spitzer IRAC 3.6~$\mu$m and 4.5~$\mu$m bands. From
Figure~\ref{fig:spectra} we see that the 3.6~$\mu$m band will quantify
the overall flux level in a continuum window.  The IRAC 4.5~$\mu$m band
will quantify the flux difference in the CO and H$_2$O absorption
feature, and should support the $\Delta F_{2.2 \mu{\rm m}}$ constraint
of a weak H$_2$O band. Because we believe that the very hottest models
are ruled out, a geometric albedo constraint from the
visible-wavelength secondary eclipse will be complementary in ruling
out cold models, especially a limit $<$~0.5 (by MOST or HST; see
Table~1).  A little further into the future, SOFIA could potentially
measure secondary eclipse closer to the planet's thermal emission peak
at $< 4~\mu$m which will help measure the $T_{eq}$.

There are many compelling questions concerning the hot Jupiters.  Do
their atmospheres have the basic composition we are assuming, such as
abundant water vapor?  Have processes such as atmospheric escape of
light gases or non-equilibrium chemistry affected the atmosphere in
ways that have not yet been considered?  Is the planet metallicity or
C/O ratio greater than solar and an indicator of planet formation
conditions?  What is the atmospheric circulation like on these planets
which exist in a radiation forcing regime unlike any planets in our
solar system---can stable supersonic winds be present?  With the
upcoming data on HD209458b and coupled radiative-transfer/atmospheric
models we can begin to answer these questions.

\acknowledgements{We thank Richard Freedman for molecular line data,
Mark Marley for useful discussions, and Scott Gaudi, Marc Kuchner, and
Joe Harrington for useful comments that improved the manuscript. This
work was in part supported by NASA Origins grant NAG5-13478 and by the
NASA Astrobiology Institute. This work is based on data taken with the
Spitzer Space Telescope, which is operated by the Jet Propulsion
Laboratory, California Institute of Technology, under NASA contract
1266284 and on data taken with the IRTF, which is operated by the
University of Hawaii under Cooperative Agreement no. NCC 5-538 with
the National Aeronautics and Space Administration, Office of Space
Science, Planetary Astronomy Program.}


\bibliography{planets}


\begin{table}

\begin{tabular}{l l l l l l}

\hline

Instrument & Wavelength ($\mu$m) & Diagnostic & Date & Comments & Ref\\

\hline

\hline

Spitzer IRAC & 3.6, 4.5, 8, 10 & flux level, H$_2$O + CO & 2005 or 06 & secondary eclipse & 1 \\

\hline

Spitzer MIPS & 24 & night side T or constraint & 2005 & primary eclipse &  2 \\

\hline

MOST & 0.3--0.8 & single band photometry & 2004 & data in hand& 3 \\
 & & geometric albedo or upper limit& & & \\
\hline

HST STIS & UV--near IR & spectrophotometry &2003 & data in hand & 4 \\
 & & geometric albedo or upper limit& & & \\

\hline

HST STIS & UV-near IR & transmission spectra &2001& data in hand & 5 \\
         &             & Rayleigh scattering, Na, K, H$_2$O &  & &   \\                  
\hline

HST NICMOS & 1-2 & transmission spectra &2004& data in hand & 6 \\
         &             & H$_2$O &  & &          \\

\hline

\end{tabular}

\caption{Upcoming observational measurements for the HD209458b atmosphere. References: (1) Spitzer IRAC GTO program PI G. Fazio; (2) Spitzer GO program 3405 PI S. Seager; (3) \citet{walk2003}; (4) HST program 9055 PI D. Charbonneau; (5) HST program 9447 PI D. Charbonneau; (6) HST program 9832 PI T. Brown.}

\end{table}

\end{document}